Journal of Risk and Financial Management

MDPI

# An Intergenerational Issue: The Equity Issues Due to Public–Private Partnerships. The Critical Aspect of the Social Discount Rate Choice for Future Generations

Abeer Al Yaqoobi [1,2] and Marcel Ausloos [1,3,*]

1. School of Business, University of Leicester, Leicester LE2 1RQ, UK; amsay1@le.ac.uk;
2. Central Bank of Oman, P.O. Box 1161, Postal Code 112, Ruwi, Oman; abeer.alyaqoobi@cbo.gov.om
3. Department of Statistics and Econometrics, Bucharest University of Economic Studies, 6 Piata Romana, 010374 Bucharest, Romania; marcel.ausloosl@csie.ase.ro
* Correspondence: ma683@le.ac.uk or marcel.ausloos@uliege.be

**Abstract:** This paper investigates the impact of Social Discount Rate (SDR) choice on intergenerational equity issues caused by Public–Private Partnerships (PPPs) projects. Indeed, more PPPs mean more debt being accumulated for future generations leading to a fiscal deficit crisis. The paper draws on how the SDR level taken today distributes societies on the Social Welfare Function (SWF). This is done by answering two sub-questions: (i) What is the risk of PPPs' debts being off-balance sheet? (ii) How do public policies, based on the envisaged SDR, position society within different ethical perspectives? The answers are obtained from a discussion of the different SDRs (applied in the UK for examples) according to the merits of the pertinent ethical theories, namely libertarian, egalitarian, utilitarian and Rawlsian. We find that public policymakers can manipulate the SDR to make PPPs looking like a better option than the traditional financing form. However, this antagonises the Value for Money principle. We also point out that public policy is not harmonised with ethical theories. We find that at present (in the UK), the SDR is somewhere between weighted utilitarian and Rawlsian societies in the trade-off curve. Alas, our study finds no evidence that the (UK) government is using a sophisticated system to keep pace with the accumulated off-balance sheet debts. Thus, the exact prediction of the final state is hardly made because of the uncertainty factor. We conclude that our study hopefully provides a good analytical framework for policymakers in order to draw on the merits of ethical theories before initiating public policies like PPPs.

## 1. Introduction

Private companies may often deal with a government in order to survive, sometimes bringing "expansion arguments" and "employment benefits". Nowadays, such deals are often subject to various (political and/or economic) conditions in order to maintain some liberal principles derived from capitalism theories. In contrast, a Public–Private Partnership (PPP) project, including the so-called Private Finance Initiative (PFI), is a government initiative (Rowland and Pollock 2002; Hodge et al. 2010), originating from the United Kingdom (UK), in 1992; it pertains to a cooperation between the government sector and the private sector in order to deliver some (in fact, public) infrastructure (Van Ham and Koppenjan 2001). Officially, the main reason for introducing a PPP is to foster the Value for Money (VfM) principle, that is, to achieve better quality services at a lower cost for the demanding public sectors (Heald 2003; Hodge and Greve 2007). The expectation is that through bundling, the whole life cycle costs of the service will be lower under a PPP/PFI scheme as compared to conventional public sector provision and finance (Engel et al. 2013). The argument is that "bundling" along with private sector provision incentivises the search for cost savings. These claims, however, are not apparently based on any formal analysis so far. Industrial economists' analyses of product bundling merely indicate that perverse incentives might arise that result in cost-cutting along with a reduction in quality (Börzel and Risse 2005).

Let it be observed that VfM assessment commences in the bidding process and throughout the project. Yet, Marques and Berg (2011) argue that there is a lack of proper regulations and measures during the bidding process of PPPs, somewhat even inciting corruption (Iossa and Martimort 2016; Schomaker 2020). In this respect, the role of the Social Discount Rate (SDR) is crucial (Roumboutsos 2010). Thus, we intend that our work contributes to the existing knowledge of PPPs schemes as well as to the considerations and subsequent implications on intergenerational equity by providing evidence of how government/policymakers manipulate SDR, whence the VfM, during the selection process of PPPS financing methods.

PPPs are essential features of capitalism (Utting and Zammit 2009); both developed and developing countries do depend on PPPs to support a country development process. Although, from the liberal perspective, cooperations between a government and the private sectors seem promising, the contracts can lead to many problems: political (less government power), economic (instability and monopoly) and social (inequality). Yet, it can be claimed that "PPPs are critically important in meeting the challenge of sustainable development" (Pattberg et al. 2012).

As well outlined by Roumboutsos (2010), "The notion of sustainability has lead to the evaluation of public projects in terms of wider socio-economic and environmental benefits. The Cost Benefit Analysis and its respective, Social Discount Rate (SDR), is of crucial importance, especially when the advantages of private financing are to be demonstrated in comparison with the alternative traditional procurement of works and services." The SDR, seen as a measure of a country's value of future costs and benefits, is related to the notion of promoted sustainability (Moore et al. 2013). The impact of smaller and declining SDRs on project selection is investigated, and a conceptual formulation concerning the selection of the project procurement method is presented. The modelled formulation will assist central and local governments in assessing projects and the potential benefit of private financing. We intend that our paper provides some further evidence of the impact of SDR used in PPPs on current and future generations well beings.

Thus, a critical research question within this sustainability framework pertains to the impact of Social Discount Rate (SDR) on intergenerational equity issues caused by Public–Private Partnership (PPP). In other words, how do public policies, based on a chosen SDR, position the society nowadays but for future generations, according to different ethical theories (or perspectives)? The answer is obtained in our paper, as we aim to propose a procedure for evaluating the SDR between pertinent ethical theories, namely libertarian, egalitarian, utilitarian and Rawlsian. We use the initiatives of the UK government as illustrating and numerical examples. (UK is considered as the front runner and initiator of PPP/PFI schemes.) Of course, we consider that the UK specificity is not a strong theoretical approximation, allowing us to imagine that the same SDR choice procedure is mutatis mutandis applicable for all other countries.

More precisely, the paper draws on how the level of SDR, taken today, distribute societies on the social welfare function and influence intergenerational issues. In doing so, the paper adds to the scant literature on the governance of PPPs by exploring the ex-post-governance issues in PPPs through the lens of incomplete contract models, rarely utilised in examining the governance of PPPs, for sustaining the wellbeing of future generations. Hence, this paper aims to catch the decision makers attention to the long-term issues caused by PPPs. Furthermore, we aim to enhance government considerations in taking precautionary measures before deciding to saddle the taxpayers and future generations with more PPPs debts.

In brief, we also consider that there are several important areas where our study makes an original contribution to the intergenerational equity field: we keep in mind throughout the study, whence focus on, the many risks that such economic and political schemes can have affecting future generations. Second, we examine the basis of practically setting the SDR in PPPs. We concentrate on two approaches put forward to determine the Social Discount Rate (SDR): (i) the social time preference rate (STPR) and (ii) the social opportunity cost (SOC). We show that the policymakers are under pressure to position themselves within the worse-off social class when they issue new policies, thereby their role and impact in choosing the proper SDR. Third, we simulate the SDR applied in the UK with respect to different ethical theories. Fourth, we provide a significant opportunity, even though at first specific, to advance the understanding of the UK government to which type of society the UK government wants to belong. Fifth, the issues discussed in this paper play an essential role in protecting the property rights of future generations.

Overall, we insist on the merits of the ethical theories that are available for the policymakers when thinking to draft public policies clearly within a moral framework in order to construct PPPs bases. For example, we emphasise the veil of ignorance of the Rawlsian theory for gearing toward a (more) equitable society. On the other hand, the policymakers can choose to increase the overall society's well beings regardless of which social class these policies affect. In such case, policymakers will draw on classic utilitarian merits. Hence, the policymakers are under pressure to position themselves within the worse-off social class when they issue new policies, the more so within the modern transparency and media communication constraints. We conclude that the policymakers today are obliged to revisit these theories' merits as guidance for forming the public policies with more considerations on the impact of these policies on the worse and better-off people.

Therefore, the paper content goes as follows: in Section 2, we develop (our) concerns about the criticality of the SDR in PPPs decision-making processes, thus endogenously introducing possible conflicts between generations. Section 3 covers the issues caused by the SDR. Section 4 highlights the parameters of social discount rates (SDR), outlining the differences between the SOC and STPR approaches used in the UK to set the SDR. Section 5 discusses the merits of four ethical theories. Section 6 discusses the impact of the SDR on intergenerational issues. Section 7 simulates the public policies between the discussed ethical theories. Our various conclusions, with recommendations for policy makers, are found in Section 8, together with mentions of our study limitations.

## 2. Intergenerational Equity Issues

*2.1. Intergenerational Equity Issues*

Intergenerational equity refers to conflicts arising between generations. Intergenerational equity issues are concerns triggered by using PPPs in delivering public services. Most of the time, these issues are ethical. This demands a compromise between current and future generations. Admittedly, current generations are privileged over future generations because they set current public policy. Indeed, any public policy endorsed today has an impact on future generations, but these generations do not have a voice in the decisions made.

One should also admit that decisions taken by the current generations should not be rationalised by either myopic behaviour or selfishness (Marini and Scaramozzino 2000; Roemer and Suzumura 2007). However, in a capitalist economy, only efficiency and allocation of scarce resources are concerned by the capital project selection process (Coyle and Sensier 2019); ethical issues are widely ignored in the cost–benefit analysis. In fact, ethical implications are neither explicit nor clearly demonstrated in economics.

The extreme level stems from Sen (1987), who argued for the separation of ethics and economics.

2.1.1. Social Discount Rate Approaches

Several approaches have been put forward to determine the Social Discount Rate (SDR). The best-known are (i) social time preference rate (STPR), (ii) social opportunity cost (SOC), (iii) weighted average approach, (iv) shadow price of capital (SPC) (Zhuang et al. 2007) and (v) the marginal cost of funds criterion (MCF) (Liu 2003). The first four approaches are widely used, while the fifth did not receive significant attention in the literature. Even though there is no consensus for which method is the most appropriate (Boardman et al. 2017), the most common paradigms are STPR and SOC (Harberger 1972; Lind et al. 2013), whence on which we focus next. The social time preference rate is the phenomenon in which people prefer consuming benefits as early as possible rather than waiting (Boardman et al. 2017). STPR reflects the "impatient" or "myopic" behaviour of people to receive money, goods or services sooner rather than later (Dasgupta and Pearce 1972). The social opportunity cost is used to measure consumption cost in a society or income accruing over time (Price 1988); it reflects the (minimum) amount the investor requires as compensation for alternative forgone investments (Harberger 1972; Zhuang et al. 2007). In a perfect market, these two rates would be the same, but due to market imperfections and externalities (e.g., taxes or asymmetric information), these rates cannot be equal (Burgess 1989; Young 2002).

Two factors taken into account in determining the SDR are efficiency and ethics. Some governments that favour off-budget financing prefer the SOC (Spackman 2017). According to Spackman, those governments weigh the social costs of public and private capital financing are nearly the same. According to Burgess and Zerbe (2011), the SOC is superior because of its generality and ease of use. The governments and authors who promoted the use of the SOC method paid less attention to the ethical concerns. Therefore, proponents of ethical issues such as climate change and all intergenerational equity issues as very long-term policy concerns support the use of STPR. As Feldstein (1964, p. 362) points out, SDR "must reflect public policy and social ethics, as well as judgment about future economic conditions". Thus, the cost–benefit analysis of long-term projects should apply STPR (Grimsey and Lewis 2005). STPR can be seen as a preferred approach for discounting PPPs, taking into account the impact on future generations. However, there are several difficulties with this. The first is that different people have different time preferences (Contreras 2014). As Grimsey and Lewis (2002) pointed out, the issue is uncertainty. Future uncertainty can result from the rate of economic growth, the amount of capital that will be accumulated, the degree of marginal utility of income of distant generations, and the state of the environment (Spackman 2004; Contreras 2014). To overcome the issue of uncertainty, Weitzman (1998), Sozou (1998) and Azfar (1999) suggested declining discount rates. In very long-term projects, the governments of the UK and France, for example, propose a declining discount rate. Both governments use Ramsey (1928) as their theoretical foundation: Ramsey formula is explained in the parameters of social discount rates section (see below in Section 4).

The death rate is one of the parameters used in the social time preference approach (STPR) to determine the SDR (see below in Section 4). According to this parameter, people prefer to spend more if they live shorter lives. Hence, they will prefer a higher SDR.

2.1.2. Temporal Choice of High or Low SDR Value

The dispute among scholars about the ethical principles of intergenerational equity is huge. The dispute dates back to seminal works by Ramsey (1928) and Pigou (1932), who proved that it is neither ethical nor irrational to use positive discounting for the current generations. Confusion arises over whether to apply either a high discount rate in favour of the current generation or a low discount rate which would allocate some of the scarce resources to future generations. One could consider that one should treat all generations equally, but what is such an equality concept? What must be equal? Too high discount rates make future benefits insignificant. They lead to underinvestment, whence to a low involvement of the public sector. On the other hand, too low discount rates lead to overinvestment, whence to an aggressive public sector.

Some scholars, such as Pearce and Turner (1990), Parfit (1992), Plater et al. (1998) and Sunstein (2009), argue that intertemporal choice between generations is insignificant and not justified. However, it is more commonly accepted nowadays that future money should be discounted due to time preference and for time value for money. In that respect, one should refer to the time preference theory of interest, also known as the agio theory of interest (Fisher 1930), in order to explain the time Value for Money (Petters and Dong 2016).

Nevertheless, the concerns of some scholars are more ethically grounded. For instance, Weitzman (1998) strongly advocated lower discount rates as an ethical consideration for future generations. Others, such as Stern (2007) and Broome (2012), also called for low discount rates. Their arguments were supported by climate change matters—the so-called global warming. Later, de Steiguer (2016) also argued that a low discount rate favours investment in future generations while a high rate is unethical because it allocates more of the scarce resources to the current generations.

On the other hand, scholars such as Mirrlees (1967) and Chakravarty (1962) claimed that existing generations might have extremely low per capita income in comparison to future generations.

Baumol (1968) also advocated a high discount rate. He argued that future generations would have several times more income than current generations. Thus, he suggested that the discount rate should be driven by the market; the public policy role is limited to the "requirements of stabilization, equilibrium in international trade" (Baumol 1968, p. 801). Instead of lowering the rate, Baumol emphasised developing durable investments. However, Milanovic (2018) argued that the original view which says "inequality means more investment and more investment means higher growth" is a myth; this is so because most of the rich capitalists either increase their own wealth without investing in society or use it for their own leisure activities.

In the same manner, Tullock (1964) claimed that an increase in society investments already means a redistribution of income from present to future. Therefore, without discounting the future generation utility, future generations become utility monsters (Nozick 1974). Grout (2003) also emphasised the use of high discount rates for PPPs. According to Grout, government procurement is less risky because the government counts for the cost it incurs to provide public services while the private sector counts for the revenue in providing public services. Although Attfield (1998, p. 207) emphasised the importance of intergenerational equity in the quality of life of future generations, he claimed that "neither future rights, nor Kantian respect for future people's autonomy, nor a contract between the generations supplies a plausible basis of obligations with regard to future generations". Accordingly, Attfield suggested "sustainable economy", "rules and policies" and "trusteeship" as reforms for intergenerational equity issues (Attfield 1998, p. 207).

In order to assess whether to apply high or low SDR, a number of measures should be taken into consideration, such as economic growth (Dasgupta and Heal 1979), uncertainty (Arrow 1995) and technological development (Kurzweil 2014). Dasgupta and Heal argued in favour of current generations while noting that the implications of future growth paths have to be taken into account. Kurzweil noted the tremendous increase of technological developments that future generations will enjoy. Arrow argued that there is an extraordinary degree of uncertainty for future generations. Deciding which generation is worse off is a very controversial matter. Although most of the economic scenarios assume that there will be greater prosperity in future, action needs to be taken today. The crucial point is whether future generations will be better or worse off; when there is a trigger for bad events, prudent measures need to be taken immediately. It is not a matter of who needs to pay for whom. The cost of preventing (or reducing) drastic (like climate-triggered) events might be less than the cost of future impacts. It is a matter of mitigating a risk that could be aggravated if left for future generations to deal with. Therefore, both setting the SDR high or low and selecting the appropriate approach to determine the SDR do matter.

## 3. Intergenerational Equity Issues Caused by PPPs

Let us turn to the fundamental question: what is the impact of SDR used in PPPs on the intergenerational equity issues?

### 3.1. Introduction

This section discusses the intergenerational equity issues caused by PPPs from the perspective of various ethical theories. The increasing number of PPPs could expose future generations to high risk through the Social Discount Rate (SDR) indeed. For example, the more resources allocated to the current generations, the more gas emission is generated by more consumption (Broome and Foley 2016). However, the consequences of such gas emission appear in the future; no reversal hysteresis is possible; this puts future generations' lives in danger. In addition to the reason given by Broome and Foley, in PPPs, it should be pointed out that appropriate SDR matters because the burden of debts accumulated today will be essentially paid by the future generations. The main risks stem from a fiscal deficit and inequitable distribution of resources. The fiscal deficit happens when a government accumulate more debts from the increasing use of PPPs. This issue allows decision makers to manipulate the SDR and make PPPs more favourable than the traditional method of financing. The inequitable distribution of resources issue arises from the government tendency to create a trade-off between efficiency and equity. The compromise between efficiency and equity distribute societies in the Social Welfare Function (SWF) according to four ethical theories, namely Libertarianism, Egalitarian, Utilitarian and Rawlsian.

These issues are addressed by answering the following questions:

A. What is the risk of PPPs' debts being off-balance sheet?
B. How do the intergenerational equity issues position the society within the SWF?

To address the above questions, a thematic analysis method is used. Practically, Braun and Clarke (2006) identify several steps to conduct thematic analysis to achieve a rich and comprehensive analysis of the studied issue. Given that, we initially got familiar with the data by reading the official transcripts multiple times. Then, we created several codes based on the intergenerational issues caused by PPPs. The study questions and sub-questions of the impact of SDR on intergenerational equity informed this process. Next, we clustered the codes based on the common issue to create initial themes. Then, the designed themes are reviewed to verify whether the extracted data and codes are fit together. Simultaneously, we considered retaining the range of initial codes while developing overarching features and higher-level sub-themes. Finally, the themes are titled, and the study was drafted.

Our work involves the usage of secondary data. Secondary data can be read from a new perspective to gain new insights (Fielding 2004). The data were mainly derived from two studies: Spackman (1991) and Pearce and Ulph (1995). Spackman lists the SDRs used in the UK from 1967 to 2003. Pearce and Ulph discuss the rationale followed by the UK in setting the SDR. Some data were also extracted from the Green Book (HM Treasury 2018). Notice that some authors are concerned about symmetry between the primary and secondary data (Long-Sutehall et al. 2010). However, the main concern of the previously used primary data is to set SDR while our study much hinges on the intergenerational equity issues caused by the SDR used in PPPs, whence rather emphasises the secondary data relevance.

*3.2. Criticality of SDR in PPPs Decision-Making Process*

SDR has a significant impact on the projects' selections. The impact becomes clear when calculating the preset values of the projects and comparing alternative methods of financing (Kossova and Sheluntcova 2016). Overestimating the value of SDR causes the dismissal of a good investment. In contrast, underestimating SDR leads to the approval of projects with long-term welfare impacts on society.

Notice how the value of this rate has a significant impact on the present value of a PPP project. An overestimated rate might lead to the rejection of a worthwhile project or shift preferences toward "more quick-impact" projects. Conversely, an underestimated rate might cause acceptance of long-term projects with presumably distant benefits to society or to rejection of private investments in government projects, or even their abandon due to unjustified otherwise political considerations (Iossa and Martimort 2016; Schomaker 2020). It is important to note that the market rate is not appropriate for discounting the benefits and costs of public projects (Grout 2003; Burgess and Zerbe 2011; Moore et al. 2013).

Moreover, the laws (or means) of regulation in project selection are more highly political matters, in our opinion, than economic or financial ones. Any selection can manifest itself as a negative realization of a submission, weakening the interconnections among the governments and private sectors in the socio-economic society network. In turn, such an action of selection makes the network itself less cohesive. Therefore, the adopted perspective does not foresee a "positive evolution" of the network—which would imply the presence of amplified effects of the selection processes on all networks.

Thus, the SDR level choice could make a PPP preferable to the traditional method of financing. While there is a perception that PPP creates VfM, this belief should be tested during the bidding process, such that PPP is compared to a traditional method. We consider that the choice of the financing method should be solely based on the VfM principle. In evaluating the VfM, the government needs to assess how much the project would cost if the government used a traditional form of finance. This is accomplished through a financial model called the Public Sector Comparator (PSC). The PSC is the public sector's risk adjusted to estimate the total cost of the project if financed by the government. A PSC assesses how much it would cost to provide the services using traditional funding. Part of the PSC is the calculation of the Net Present Value (NPV), where the SDR is used to discount future cash flows. PSC is a hypothetical costing and not to be confused with a genuine government bid because PSC also includes the risks of the project. The PSC is complex, consisting of a detailed timetable of works and a series of costs. The PSC compares the project's NPV as if it were wholly financed by the government to the NPV if the project were financed by the private sector through a PPP. The selected method of finance depends on which of the two has the highest NPV. However, decision makers can still bias the results towards their preferred method of financing by adjusting the level of SDR used in the PSC to make it appear more expensive than a PPP. The criticality of the level of SDR is demonstrated in the following example.

In the Carlisle Hospital PFI scheme, the actual cash cost of the PFI against the PSC was GBP 577 million and GBP 550 million, respectively (Rowland and Pollock 2002). That shows that the PSC bid was GBP 27 million cheaper than the PFI bid. Nevertheless, the PFI scheme was selected for the project because of the discount rate used in calculating the NPV. The NPV was calculated as GBP 173.1 million for the PFI and GBP 174.3 million for the PSC. That made the PSC appear GBP 1.2 million more expensive than the PFI. The NPV demonstrates that deferred payment is better than an immediate payment. Table 1 shows that the option of PFI would not be selected if the discount rate was altered to 5.5%, i.e., 0.5% less than the applied rate.

Table 1 reflects the SDR's implications on the selection process. Table 1 shows that PFI would be dismissed if the SDR was not used to discount future cash flows. Table 1 also shows that using lower discount rates can make traditional financing preferable. Yet the decision maker may keep increasing the SDR to 6% to make PFI lower than PSC. In contrast, a slight reduction in the SDR by only 0.5% makes PFI look more expensive.

**Table 1.** Comparison between Net Present Value (NPV) of Private Finance Initiative (PFI), scheme and Public Sector Comparator (PSC) using different discount rates in the case of Carlisle Hospital (figures in GBP millions (m)). Source: Rowland and Pollock (2002, p. 23).

| Discount Rate (%) | PFI | PSC | Difference in Favour of PFI |
|---|---|---|---|
| 6 | GBP 173.1 m | GBP 174.3 m | GBP 1.2 m |
| 5.5 | GBP 186.7 m | GBP 185.8 m | − GBP 0.9 m |
| 5 | GBP 202.0 m | GBP 198.8 m | − GBP 3.2 m |
| 4.5 | GBP 219.5 m | GBP 213.9 m | − GBP 5.6 m |
| 4 | GBP 239.3 m | GBP 231.2 m | − GBP 8.1 m |
| 3 | GBP 288.6 m | GBP 275.0 m | − GBP 13.6 m |

Therefore, an SDR appropriate choice in long-term projects matters for several reasons. First, SDR determines the present value of future payments and how fast commodities' value diminishes over time. An example is given by Broome (2008) to compare the high Nordhaus (1994) 6% SDR and Stern (2007) 1.4% SDR. As a result, USD 1 trillion in 100 years is worth USD 2.5 billion today using the Nordhaus (1994) discount rate, while for the same period worth it is worth 247 billion dollars using Stern (2007) SDR. In addition, Howarth (2005) argued that people consider the uncertainty factor in discounting and the "productivity" of the invested capital.

Hence, the selection of the appropriate SDR is critical in deciding the financing method. The partiality in adjusting the level of SDR to favour the PPP is justified by Grout (1997), who states that a PPP is the only option available for the government. Another example of the crucial importance of SDR in the selection of financing method is the Haringey school scheme (Rowland and Pollock 2002). In this scheme, the discount rate was changed several times to make the PPP appears to be the cheaper option. Thus, the partiality of SDR selection destroys the purpose of the VfM, which is the core of PPPs.

*3.3. Issues Caused by the SDR Choice*

PPP contracts can last many years, so decisions made today will have consequences for future generations. The main two consequences are future fiscal deficit and equitable resource distribution between generations.

3.3.1. The Risk of PPP Debt on the (Long-Term) Fiscal Deficit

The fiscal decisions of taking more on PPP liabilities are a burden that will be borne by future generations. Theoretically, PPPs should cut costs by bundling services and improving infrastructure quality by using private sector expertise. However, some authors argued that PPPs are a mechanism to bypass government budget constraints (Grout 1997; Smith 1997; Schwartz et al. 2008). The off-balance sheet makes these liabilities not available for public scrutiny, which allows the government to accumulate more loans. Parts of these liabilities are the guarantees reported within the contingent liabilities. The accelerated rate of these liabilities will be paid by future generations, either through taxes or cuts in expenditure (Regan et al. 2011).

PPP debt and the guarantees provided by the government to the private sector are off-balance-sheet items. They are not parts of state debt statistics. This mitigates the risk that global credit rating agencies will downgrade the government's credit rating. Off-balance-sheet items facilitate the accumulation of debt that will have to be paid by future generations. Guarantees given in PPPs are meant to mitigate systemic risks, such as environmental and political risks. These guarantees are reported as off-balance-sheet items because they are contingent on the occurrence of the risks agreed on. Most developing countries, especially those with a low credit rating, offer such assurances to attract investors. The guarantees can take the form of income support, such as "loan guarantees, deposit insurance … and trade and exchange rate guarantees" (Towe 1991, p. 111). The issue is that off-balance-sheet debt and guarantees are not accounted for in the debt-to-GDP ratio.

A high debt-to-GDP ratio can expose the government to a severe fiscal crisis. In the UK, the total credit to the private sector non-financial sector as a percentage of GDP ratio ranged between 185.1% (Q4; 2009) and 155.2% (Q1; 2019); see FRED (2021). This is higher than the triggering ratio of total debt-to-GDP of 150% (Keen 2017), for which Keen "expects" that such a high debt-to-GDP ratio will endogenously trigger the "next" financial crisis (Shmelev and Ayres 2021). Countries with a debt-to-GDP ratio higher than 150% are considered to be in the danger zone. Although the ratio of PPP debt to GDP is approximately 1.5% in the UK (World Bank 2020), this ratio incorporates only PPPs held by the central government. The UK government excluded the local authorities' PPPs debt from the 1.5% as the central government will bear the risk. Hence, the real PPP debt to GDP ratio is higher than the reported 1.5%. However, to monitor the real PPPs debt to GDP ratio in the UK, the data have to be accurate. They need to be monitored and observed closely to maximise the total surplus generated from PPPs.

To control off-balance-sheet items, the government needs effective systems. The current systems account for contingent liabilities when the actual obligation costs are paid (Towe 1991) rather than when the contingent liabilities are incurred, following cash basis accounting rather than not accrual basis. The current systems used by some government departments are identified by Towe (1991). The first is the System of National Accounts (SNA). The SNA is "designed to integrate balance sheet information with information on production, income, savings and real and financial investment" (McCulla et al. 2015, p. 2). However, the SNA neglects the fiscal risk of PPP debt obligations (Schwartz et al. 2008). The other system is Government Finance Statistics (GFS). GFS contains detailed data about the country's "revenue, expense, transactions in assets and liabilities and stock positions in assets and liabilities of general government and its subsectors" (IMF 2020). GFS compiles the data on both a cash basis and an accrual basis. Accrual helps decision makers to trace accumulated government debt. Although the GFS use a cash and accrual basis, it only covers the on-balance sheet items. For example, in Oman, the Ministry of Finance lacked a system to measure the impact of aggregate debt (on-balance-sheet and off-balance-sheet obligations) on the long-term fiscal deficit. Hence, policymakers might decide to increase the number of PPPs without taking precautionary measures.

3.3.2. Justice of Resource Distribution between Generations

Decisions on how to allocate resources between generations are difficult because of the trade-off between efficiency and equity. Should policies increase the allocation of resources on an efficiency basis or on an equitable basis? Or should policies be impartial and treat all generations the same, regardless of who will be worse off? Efficiency is achieved when maximum use is made of resources. Efficiency depends on the opportunity cost of capital; the higher the opportunity cost, the more efficient the allocation. The more efficient the market, the more freedom individuals have and the more they use their power and compete to acquire from the available resources. There is less government interference in an efficient market. As a result, people who have power, authority and easy access to resources will reap as much as they can from the market.

In such a market, a high SDR should be set. A high SDR promotes economic growth at the expense of inequitable resource distribution among people or generations. In the UK, the high SDR before 2003 reflects the high growth in that period. Yet, it also proves that SDR reduction after 2003 was not based on equitable distribution of resources between generations. For example, in the UK, the Gini coefficient of total wealth was 63% for the period from 2016 to 2018 (ONS 2019). Physical inequality was 47% in the same period, while net property and pension wealth accounted for 66% and 72%, respectively. According to the ONS (2019), from 2016 to 2018, almost 45% of the total wealth in the UK was controlled by just 10% of the population. This ratio has remained approximately the same since 2006–2008.

The high inequality in the UK is reflected in the discrepancy between the low SDR of 3.5% used in the country since 2003 and the UK's high Gini coefficient. Although a low SDR should reflect a low level of inequality between social classes, the statistics in the previous paragraph show the opposite. The discrepancy between the SDR and the Gini coefficient proves that the reduced SDR was not equitably set. This is mentioned by Pearce and Ulph (1995), who argues that the 6–8% rate was too high; even if it is difficult to approve any appropriate SDR choice, it can at least be reduced for climate change matters.

## 4. Parameters of Social Discount Rates (SDR)

There are several different ways to determine the SDR, including STPR, SOC, weighted average approach and SPC (Zhuang et al. 2007). The choice of approach is governed by the project's impact on the country's local spending, the cost of global borrowing, and private investment (Zhuang et al. 2007). There is no consensus on a preferred approach, but the SOC and the STPR are widely used.

The following section provides an overview of the parameters used in the two approaches. The focus will be on the STPR because it is the UK's current approach used. The SOC will also be discussed because it was used in the UK before 2003. The SDR used can be real, adjusted for inflation or nominal (Zhuang et al. 2007). The "real version" is used in the UK.

### 4.1. Social Time Preference Rate (STPR)

The STPR approach takes into account certain social aspects. These aspects are individuals' preferences, level of consumption and saving, and growth rate. STPR has been followed in the UK since 2003. It estimates how current generations can defer current consumption in order to consume more in the future (HM Treasury 2018; Zhuang et al. 2007). The parameters underlying the STPR approach are supposed to consider people preferences and the marginal utility of consumption, but it is argued that this is unmeasurable (Pearce and Ulph 1995). Hence, "the after-tax rate of return on government bonds or other low-risk marketable securities" is used to determine this rate, with some consideration given to people preferences (Zhuang et al. 2007, p. 4). This distinguishes this approach from the SOC, which is merely based on efficiency.

The most commonly used formula to determine the STPR is Ramsey's formula (1928). Ramsey's formula combines the utility discount rate (which reflects the pure time preference $\varrho$, the elasticity of the marginal utility of consumption $\mu$ and the annual growth rate of real consumption per capita $g$ (HM Treasury 2018):

$$r = \varrho + \mu\, g \tag{1}$$

where *r* means STPR here above.

In the UK, $r = 3.5\%$: $\varrho \approx 1.5\%$; $\mu = 1.0$ and $g = 2\%$.

(a) The time preference ($\varrho$)

The time preference (ϱ) is depicted in the following formula:

$$\varrho = \delta + L \qquad (2)$$

The first term δ is the allowance for the pure time preference. This rate "discount[s] the welfare arising to people in the future purely by virtue of this utility arising later" (Pearce and Ulph 1995). The second term L is any unforeseeable or systemic risk that may occur in the future. For example, "rapid technological advances that lead to obsolescence, or natural disasters that are not directly connected to the appraisal" (HM Treasury 2018, p. 102). However, Pearce and Ulph (1995) argue that L denotes changing life chances over time. The justification of L as life chances is derived from the STPR definition mentioned above. This justification demonstrates that people's preference for getting a return on their investment today or in the future hinges on their expected length of life. This rate L is computed by dividing total deaths by total population. For the UK, Spackman (1991) calculated this to be 1.1%, while the "Green Book" (HM Treasury 2018) estimated the value of L to be 1%, with δ = 0.5%, such that ϱ = 1.5%.

This value is greater than the time preference rate suggested by Stern (2007); he suggested a pure time preference of δ = 0, which took into consideration intergenerational long-term ethical issues such as climate change. He also estimated the risk parameter L to be 0.1%. Stern interpreted this parameter from a social welfare perspective, while the Green Book included other risks. The Green Book suggests that the risk parameter L should include either environmental or "man-made" risks. The main difference between the Green Book and Stern is the Green Book adopted a short-term horizon when estimating the time preference rate while Stern was concerned about long-term ethical matters. The Green Book suggested separate cost–benefit analysis of the intergenerational issues without incorporating them in the computation of the SDR.

Notice that Freeman et al. (2018) specified that the allowance for pure time preference δ should be between 0 and 1%. The 0–1% logic is shown by Broome: "A universal point of view must be impartial about time, and impartiality about time means that no time can count differently from any other. In overall good, judged from a universal point of view, good at one time cannot count differently from good at another. Therefore, ... the [pure time] discount rate ... must be nought" (Broome 1992, p. 92).

The zero per cent allowance means being impartial. Impartial means giving the same weight to worse-off and better-off generations regardless of their interests, yet treating everyone the same means giving more weight to the better-off generation when the expectations show high future economic growth (Pearce and Ulph 1995). This approach also can raise ethical issues because a positive, pure time preference rate gives a clear indication that future generations are better off, which might not be the case. Pearce and Ulph (1995) suggest that any appropriate pure time preference rate is hard to predict.

(b) The elasticity of marginal utility (µ)

The elasticity of marginal utility µ of consumption measures the relationship between the reduction in marginal utility for every increase in consumption. The rate of marginal utility can be measured by observing either how people save or) by how "worthwhile it is to carry out transfer of income from a rich person to a poor person depending on how well-off the two are" (Pearce and Ulph 1995, p. 12). Groom and Maddison (2018) believe that regardless of approach, the more convincing elasticity of marginal utility is 1.5%. The Green Book applied an elasticity of marginal utility of 1%, as suggested by Stern (2007) on an intergenerational basis. Groom and Maddison (2018) argue that Stern's figure was too low an estimate of the degree of inequality aversion. Yet the rate of 1% is widely accepted in several countries that follow the STPR approach, such as the UK, Malaysia and Japan.

(c) The growth rate estimation (*g*)

The Green Book sets the growth rate *g* at 2%. It was determined from the real consumption growth of the post-war period, 1949–1998. Groom and Maddison (2018) estimated a growth rate of 1.1%. Their estimation included the period from 1830 to 2009. The low growth rate found by Groom and Maddison was not affected by data smoothing. According to them, SDR "differs depending on whether smoothed or unsmoothed data is used" (Groom and Maddison 2018, p. 1174).

Recall that the STPR is mainly concerned with the SOC of forgone consumption. Hence, the STPR is lower than the SOC approach. However, Zhuang et al. (2007) argue that the STPR is low because it allows the government to choose the low-return public investments. Zhuang et al. are more concerned about efficiency, while Pearce and Ulph (1995) are concerned about climate change. Regardless of which approach is better, the government sets the approach and the level of SDR in a trade-off between efficiency and equity.

*4.2. Marginal Social Opportunity Cost (SOC) of Capital*

The marginal social opportunity cost (SOC) of capital is based on the competition between the private sector and the government sector for the same resources. As Zhuang et al. (2007) put it, "government and private sector compete for the same pool of funds" (p. 9). In this case, the government will seek a high return to offset the forgone fund of the private sector. If not, then the government should account for the forgone SOC by displacing the private investment. In both cases, if the government invests for a high return or accounts for the opportunity cost of forgone private sector investments, the SDR will be high.

The SOC can be determined using the pre-tax rate of return on safe, private investments (Zhuang et al. 2007) of the top-rated corporate bonds (Moore et al. 2004). The pre-tax rate of return can be computed using the marginal pre-tax rate of return instead of the average rate because when investors evaluate the viability of any project, they make their estimation from the marginal rate of return. Second, the rate of return of private investment includes a risk premium that compensates the investors for the risk they take from the investment. The public investment has more tolerance for the risk; hence, it makes sense to use the marginal rate of return. Yet all these considerations do not take into account an equitable distribution among people within a single generation or between generations. The weight given to such considerations in the SOC approach is zero.

By following the SOC approach, the opportunity cost of capital will be higher than in traditional government financing. For 1991–2003, the UK used the real discount rate of 6% (Blaiklock 2011). If the same rate were adjusted for inflation, it would be an 8–9% nominal interest rate. In the same period, the cost of capital for a 30-year UK bond had an average nominal interest rate of 5.5–6%. That shows a difference of 2.5–3%; the PPPs cost is higher in comparison to the conventional financing if it is adjusted to the nominal rate of interest. By 2003, the UK government had applied a real interest rate of 3.5% following the STRP approach, which was much less than 6%.

*4.3. STPR and SOC from an Ethical Perspective*

The parameters of the social time preference rate (STPR) are meant to bring balance in the trade-off between efficiency and equity. It pulls the curve in Figure 1 down towards equity. Unlike the STPR, the social opportunity cost (SOC) is concerned with distributing resources between the private sector and the government sector in an efficient way. In brief, the SOC supports more efficiency, which is the main drive of libertarianism. Let us specifically consider in more detail both perspectives.

Of the three parameters in STPR, the equity consideration is found in the estimation of the elasticity of marginal utility of consumption µ. In this parameter, a redistribution from the high-income to the low-income group was applied in the determination of (µ).

The value μ = 1.5, currently used in the determination of the UK SDR, "implies that the higher income group's extra £1 is valued at only 35% of £1 to the lower income group" (Pearce and Ulph 1995, p. 16). Pearce and Ulph (1995) argue that this treatment is egalitarian because it tries to equalise people. In fact, this is only one method of determining μ. Other methods treat all consumption of better-off and worse-off people as the same. Usually, μ is computed by calculating the total amount of consumption in a society. This method is followed in the UK; as the consumption used in the SDR of the UK is an aggregation of total consumption, its value/choice does not differentiate between rich and poor.

The advocates of the low discount rate attribute the reason to the climate change matters and other problems future generations might face, such as "long-term welfare concerns e.g., radioactive waste disposal" (Pearce and Ulph 1995, p. 6). The change in SDR calculation introduced in the UK in 2003 stemmed from two reasons. First, the 6% used before 2003 was too high (Pearce and Ulph 1995). Second, the STPR approach included social parameters such as individual time preferences.

On the other hand, SOC is a good option as SDR when there is competition between the government and private sectors over the resources (Kossova and Sheluntcova 2016). Moreover, to say the least, such a competition may not necessarily motivate the private sector to provide reasonable cost and better quality unless it is not a monopoly market like PPPs cases. Yet, we argue that if the project's objective is to offer social benefits to society, as in the case of PPPs, then STPR is a better option.

Zhuang et al. (2007) recorded the SDRs of different countries for that time. Developing countries such as India, Pakistan and the Philippines followed the SOC approach to measure the SDR. Hence, these countries SDRs are high. The SDRs are 12% in India and Pakistan and 15% in the Philippines. (Zhuang et al. attribute the high SDR in developing countries to several reasons: lack of access to the international markets and low financial resources. Hence, the intergenerational matters in these countries are not really a matter of concern). In contrast, some developed countries such as Italy and Spain also apply "high" SDR of 5% and 6%, respectively, following the STPR approach; Norway's SDR is 3.5%; it is driven by the real rate of government borrowing; Germany's SDR is 3% based on the federal real refinancing rate.

**5. Ethical Theories Related to Intergenerational Equity**

*5.1. Merits of Libertarianism*

Libertarianism is one of the philosophical theories that call for unimpeded freedom and autonomy. The theory ranges from absolute freedom of the individual in all life aspects to more organised freedom. Absolute freedom allows the power of the market to organise relationships between people and the private and government sectors. Market autonomy requires no regulation and minimal state interference. At the other end of the libertarian scale, the state interferes in some aspects, such as political and legal systems, but it calls for laissez-faire in terms of organising the economic system.

According to Freeman and Phillips (2002), libertarianism is a thrust for irresponsibility toward others. Freeman and Philips set the following attributes for the libertarian: (1) relies on freedom, liberty, the equal liberty principle, or some kindred notion; (2) relies on basic negative rights, including individual property rights; (3) allows for the creation of positive obligations through various voluntary actions (e.g., contracting and promising); (4) countenances at most a minimal state; (5) assumes that human beings are largely responsible for the effects of their actions on others (Freeman and Phillips 2002, p. 336).

Libertarianism and capitalism are two tangled ideologies. Libertarianism is a political system, while capitalism is an economic system. As mentioned above, libertarianism gives infinite property rights to individuals. On the other hand, capitalism gives infinite rights to the private sector. Although theoretically, there might be a difference between them, in practice, they are entangled. The property rights of the private sector are owned by individuals. These individuals own the resources in the capitalist markets. Consequently, this allows them to possess more property rights in liberal markets. The entanglement between the economic and political systems creates regulatory capture, which combines wealth and power. Given that, the more wealth and power you have, the more scarce resources you possess. This can justify the SOC approach used in the UK from the 1960s to the 1990s.

*5.2. Merits of an Egalitarian Society*

Egalitarianism, in theory, is a policy that attempts to equalise the utility level of all individuals. It prefers equality to equity in utility distribution. It does not matter whether this equality causes misery to be shared by all individuals. Adherents of egalitarianism prefer a situation where all society members possess less total utility and less wellbeing than where the utility of all individuals varies with a much higher total wellbeing in the society. Thus, this theory can lead to poverty and shirking behaviour in a society. Yet, this eliminates the motivation to work or to make any extra effort to increase individual wellbeing because each individual already owns the same amount of utility. In an egalitarian society, the state government officially forces redistribution of wealth through the tax system or transfer systems.

Egalitarianism gives all people the same property rights even if that produces a smaller total surplus. In the context of the SDR, the SOC approach is outside the scope of egalitarian theory because there is no equality in SOC. However, the ($\mu$) parameter in the STPR redistributes resources between different classes. This parameter can be computed through different methods, one of which is the thought experiment. This method redistributes resources from high-income to low-income people (Pearce and Ulph 1995). Pearce and Ulph describe such processes as "strongly egalitarian" (p. 16). In this method, more weight is given to low-income people by transferring resources to them from high-income people for the sake of equality rather than equity.

Advocates of egalitarianism will not apply equally to the extreme where the state will take resources from someone to make everyone in a society exactly equal. Advocates of egalitarianism support the equality of everyone possessing the same resources and opportunities at the start of life. Individuals should be entitled to keep what they earn through the choices they make and the effort they exert. Resources not acquired through effort and open (fair) opportunities will be redistributed within society. Egalitarians believe that how well-off people are should be determined only by the responsible choices people make and not by differences in their circumstances that they did not choose. Arneson (2013) criticises egalitarianism for being a complicated and difficult task of distinguishing between fair opportunities and the fruit of luck. It is also difficult to measure the effort people make to reflect whether they deserve their possession. Moreover, moderate egalitarianism—giving the same opportunities to all generations—might suggest an SDR of zero. However, even if the moderate view of egalitarianism is followed in public policy, it is still difficult to assume that zero SDR is impartial, as Broome (1992) suggests, because, in intergenerational equity issues, worse-off people also matter.

*5.3. Merits of a Utilitarian Society*

Utilitarianism is a teleological theory that concerns the consequences of any action on the total wellbeing of society. From the utilitarian perspective, the right action is the one that produces the most happiness for the benefit of society, regardless of who receives it. Utilitarianism simply states that an act that does the greatest good for the greatest number of people is generally good. "Good" in this sense can be in the form of satisfaction, pleasure or happiness. Total social wellbeing will increase if the utility of any individual increases and vice versa (Roemer and Suzumura 2007). In the utilitarian view, the redistribution of wealth matters only if it affects the total value or total results it produces.

The theory is categorised into two opposing strands: act utilitarianism and rule utilitarianism. Act utilitarianism is intuitive and judgemental, while rule utilitarianism is rigid and idealistic. Act utilitarianism depends on the consequences of the action taken. Whether an action is good or bad is decided by its results. An action is considered good if it benefits or makes the majority of people happy even if it causes severe harm to a minority of people. On the other hand, rule utilitarianism is all about following the rules. It is believed that setting rules maximises the utility for the majority. Because even if following the rule will cause severe suffering for a minority, the rule is intended to cause great happiness for the majority. Accordingly, even if the intention of the action or rule is not moral, if that particular action results in providing greater happiness for others, then it is ethical and morally right under utilitarianism.

This theory is identified as encouraging the compromise of individual interests while also focusing on promoting a greater sense of common good in society. It does not matter where this utility is coming from as long as the total utility of society is maximised. Other than maximising the total wellbeing of people, utilitarianism introduced the Diminishing Marginal Utility (DMU) principle.

According to Bentham, the originator of utilitarian theory (Postema 2006, or see https://plato.stanford.edu/entries/bentham/#LifWri; accessed on 28/02/2020), the DMU principle means the more commodities people possess, the less impact a marginal utility will have on their total utility. In other words, the more commodities people possess, the less utility they will enjoy from a marginal increase in these commodities. Bentham argues that rich people can give away some of the commodities they possess to poor people. Transferring GBP 100 from a rich person to a poor one will not affect the total utility of the rich person, but it will cause a huge increase in the utility of the poor person.

A hybrid between egalitarianism and utilitarianism is weighted utilitarianism. In weighted utilitarianism, policymakers should aim to maximise the aggregate utility but assign greater importance to less well-off people. This means that the aggregate social welfare would be maximised but to the level that will not make worse-off worse.

*5.4. Merits of a Rawlsian Society*

Rawlsian theory, or "justice as fairness", was founded by a philosopher (Rawls 1971). Rawls's theory consists of two principles: (i) the "liberty principle" and (ii) the "difference principle". The liberty principle states that basic liberties have to be secured for all people. It promotes the idea of having as much liberty as possible. However, each person's liberty should not infringe other individuals' liberty. The basic liberties promoted by this principle are: (i) political liberty (the right to vote and to hold public office); (ii) freedom of speech and assembly; (iii) liberty of conscience and freedom of thought; (iv) freedom of the person, which includes freedom from psychological oppression and physical assault and dismemberment along with the right to hold (personal) property; and (v) freedom from arbitrary arrest and seizure (Rawls 1971, p. 53).

The difference principle has two parts: (i) society should stop redistribution when redistribution worsens the situation of the least well-off individual and (ii) open access for all. In the first part, Rawls argues that there should be a distribution that makes everyone better off, but if not, equal distribution is the optimal choice. In other words, unless the utility of one person maximises the utility of another person, the wellbeing of society as a whole will not increase. This principle has implications for taxation and redistribution policies. The other part of the difference principle is equal opportunities. This part concerns providing open access for all individuals to "positions of authority and responsibility" (Rawls 1971, p. 53). However, this principle has two limitations. The first one of all people having to enjoy equal basic rights, which are equal basic liberties, must be fully satisfied, and institutions ensuring fair equality of opportunity must be in place.

Rawls claims that people will be reluctant to take risks beyond certain limits. In Rawls's theory, this principle is known as the "veil of ignorance". It assumes that policymakers might belong to any of the social classes in society. The chance of being in the better-off class is the same as being in the worse-off class. Therefore, policymakers will set policy impartially. They will also be very likely to evaluate the impact of their policies on the worse-off group because they might belong to it. Given that, the worse off people will be given high weights during the decision-making process.

Rawls advocates the moral psychology of reciprocity. In a justification of reciprocity, the theory assumes that people "answer in kind" or repay kindness by kindness (Rawls 1971, p. 433). For this reason, Rawls assumes his theory is more resilient than utilitarianism because utilitarianism is developed from the moral psychology of sympathy, while Rawls's theory is based on the moral psychology of reciprocity. Rawls believes that people have the tendency to put aside their own interests in order to promote fairness and justice for other people. This tendency, according to Rawls, is found in infants in its basic form, and it becomes woven into their development of morals and ethics. These morals, under favourable conditions, guide their decisions and actions when they are mature.

Rawls's theory is criticised because the reciprocity principle and veil of ignorance are not realistic (Wenar 2017). The reciprocity principle is not convincing because people will not be incentivised to compromise unless they are assured that others will also do the same. Moreover, the veil of ignorance principle can be challenging because the intention of the policymakers cannot be ascertained. That will greatly depend on the policymakers' beliefs and values.

*5.5. Trade-Off between STPR and SOC within the Ethical Theories*

There is consensus among economists, including Spackman (1991) and Pearce and Ulph (1995), that the SDR is judgemental and there is no appropriate level of SDR. The approach used to determine the SDR can also vary according to what kind of society the government wishes to create. For example, a libertarian society is motivated by efficiency, which involves a free market and competition. Burgess and Zerbe (2011) argument for recommending the SOC as the most appropriate approach is valid. The SOC is easy to use, with a single direct rate. A libertarian society might apply the market interest rate as compensation for the forgone private sector investment. The SOC also can be used in the classic utilitarian model, where the aim is to maximise the total surplus consistent with current practice in GDP computation.

However, an egalitarian society might apply the STPR approach with adjustments for better-off and worse-off generations. In an egalitarian society, the parameter of time preference ($\varrho$) might not be considered because all generations should be treated the same and people preference does not matter. Likewise, the time preference parameter ($\varrho$) might be given a low weight in the weighted utilitarian and Rawlsian approaches because the focus of these two theories is the worse-off group rather than people's preferences. More weight in these theories will be assigned to the elasticity of the marginal utility of consumption ($\mu$) and the annual rate of growth per capita real consumption ($g$) to determine the worse-off generation.

**6. Sensitivity of Intergenerational Matters to the Level of SDR**

There is no consensus among decision makers and economists on the appropriate SDR. Stern (2007) provides arguments that support future generations. He suggests that a 1.4% SDR would protect the interests of future generations. Stern's stance is based on his ethical values; he believes in making available resources more sustainable. On the other hand, Nordhaus (1994) claims that there is no urgent need to save resources for the future. Nordhaus suggests a discount rate of 6%. He gives less weight to future generations. Other scholars, such as Pearce and Turner (1990), Parfit (1992), and Plater et al. (1998), suggest using an SDR equal to zero.

Thus, since the chosen rate depends on the values of the decision makers who set the SDR, economists have identified factors to help decide which generation should consume more resources; this is discussed next.

In ethical economic theories, as discussed in Section 5, better-off generations have to get less from scarce resources. The compromise between generations and the question of who should bear the costs depend on three main indicators: economic growth, technological evolution and event uncertainty.

*6.1. Economic Growth*

Economic growth is used as an indicator of future prosperity. Studies expect that economic growth will continue in the future, and the standard of living will be much better than today (Dasgupta and Heal 1979). The OECD (Johansson et al. 2012) expects the world economic growth rate to surpass an average of 3% annually up to 2060. Likewise, the OECD predicts that GDP will increase on average by 2% per annum up to 2060. Pearce and Ulph (1995) claim that the growth in "capital accumulation and technical change" can make descendants better off than current generations. Therefore, provided that future generations are better off, a high SDR in favour of the current generations is defensible. Assigning a high SDR to worse-off generations is the crux of Rawlsianism.

However, high economic growth can also indicate that future generations are more productive. Pearce and Ulph (1995) argue that in the classic utilitarian model, more weight would be given to more productive people to increase the total surplus of society. The assumption of continuing economic growth means that descendants deserve more resources. The classic utilitarian point of view indicates that low SDR for the current generation is defensible. However, it is not equitable because it means worse-off generations might suffer more for the sake of efficiency. This premise supports capitalism. (The optimal point in Figure 1 shifts toward a more libertarian society; see more discussion below.)

The above scenarios can also be a foundation for the government (the owner of the property rights) to decide on the appropriate SDR. The SDR chosen is reflected in the public policies declared by the government. An equitable government will set public policy based on Rawlsian principles, and more weight will be given to the generation which has less economic growth. On the other hand, the efficient government will assign more weight to productive generations creating a utilitarian society.

*6.2. Technological Evolution*

Technology is expected to be more developed in the future than it is today. Kurzweil (2014) defends the prosperity of future technological developments. Kurzweil argues that "period during which the pace of technological change will be so rapid, its impact so deep, that human life will be irreversibly transformed" (Kurzweil 2014, p. 393). Although economic growth and uncertainty indications make the future looks vulnerable, technological development could make life in the future much better. A decade ago, no one expected digital technology to invade people's lives in the way that it has today. This could justify the current generation thought that future generations would be better off. However, this indicator cannot be used on its own, and its ethical values cannot be quantified.

*6.3. Event Uncertainty*

Uncertainty is a vital factor in predicting the future; better circumstances for future generations cannot be assumed. High uncertainty about the future makes the distribution of resources between current and future generations more complicated. The dilemma is that even if there were positive economic growth and rapid technological change in the future, extraordinary (rare) events might occur. As usually admitted, human bounded rationality cannot predict all future contingencies (Simon 2000), such as wars, environmental disasters and the threat of security issues (Broome 1992). Uncertainty makes differentiating between better-off and worse-off generations more complicated.

To complicate matters further, the effects of climate change, such as heatwaves, storms and floods, are all expected to increase. In 2003, almost 35,000 people died in Europe as a result of high temperatures. The World Health Organization estimated that about 150,000 people had already been adversely affected by climate change by 2000 (Broome 2008). However, these numbers might decrease as a result of preventive measures the governments might take. Healthier lifestyles may also increase life expectancy in future generations, but this will not prevent the effects of climate change. Therefore, economists such as Pearce and Ulph (1995) argue that regardless of who will be better off and worse off, climate change requires consideration. Climate change matters happening today have an impact on future generations.

To protect future generations, Rawls (1971, p. 252) emphasises that "each generation must…put…aside in each period of time a suitable amount of real capital accumulation". Capital accumulation can take different forms. One form is a long-term investment, including capital investment or financial investments. Another form of investment is in the "stock of skills and productive knowledge embodied in people" (Harrison 2010, p. 3). Harrison suggests investment in laws and regulations; in other words, in human capital and institutional capital. Human capital would support the growth of the country if coupled with robust regulation. Such measures would create better opportunities for current generations while still protecting the rights of future generations.

The three indicators discussed above are subject to debate. Kula (1985, 1987) argues that predicting the future is a formidable task and that the three indicators are not good measures to decide which generation is better off. Whether or not future generations will be better off, the advantage held by current generations is that they have the right to negotiate or abandon the unacceptable government policies. Today's decisions have an impact on the future generations, which are not able to bargain or remind decision makers of their rights. Current generations might prefer to receive as much as possible today without paying attention to the future generations. The ones who are supposed to consider future generations in their decision are decision makers. Decision makers might prefer efficiency over equity, especially in capitalist markets, where ethical issues are likely to be neglected. Yet, decision makers might still seek a balance between protecting the property rights of future generations and efficiency by simulating the SDR between the ethical theories (see next Section 7).

## 7. The Position of the State on the Social Welfare Function Based on the Used SDR

In the following section, the different public policies are analysed using a number of ethical theories in view of sustaining our conclusion for future generation sustainability. The policies used in the UK between 1967 and 2003 are commented upon.

### 7.1. The Social Welfare Function (SWF)

The graph presented in Figure 1 provides essential support for the discussion in this paper. The figure shows the positions of four ethical theories—libertarianism, egalitarianism, utilitarianism and Rawlsianism—on the economics of welfare. These theories represent the ethical distribution of wealth and income in society. In a liberal society, autonomy is high, and there is less government interference in increasing welfare and wealth distribution. The tax and transfer system is equally applied to all social classes. The aggregate wellbeing of such a society increases by efficiency. Thus, a liberal society goes at the top of the SWF graph.

In an egalitarian society, everyone is treated exactly the same. The "egalitarian equilibrium line" divides the graph into two identical parts; see Figure 1. At that point, efficiency is equal to equity, and the trade-off between them is zero. An egalitarian society is not productive because it promotes equality and totally neglects equity. It encourages people to shirk because they get the same property rights as productive people. The classic utilitarian society is more moderate than the egalitarian. It seeks equality that increases the total aggregate wellbeing of the society regardless of whether worse-off people will be even worse off.

In neoclassical utilitarianism (weighted utilitarianism), increasing total society wellbeing is an objective, but high weight is given to worse-off people. Weighted utilitarianism is located between egalitarianism and Rawlsianism in the SWF. The Rawlsian society focuses more on equity. It concerns the wellbeing of the worse-off group. Therefore, it is closer to the equity side of the SWF in the trade-off graph.

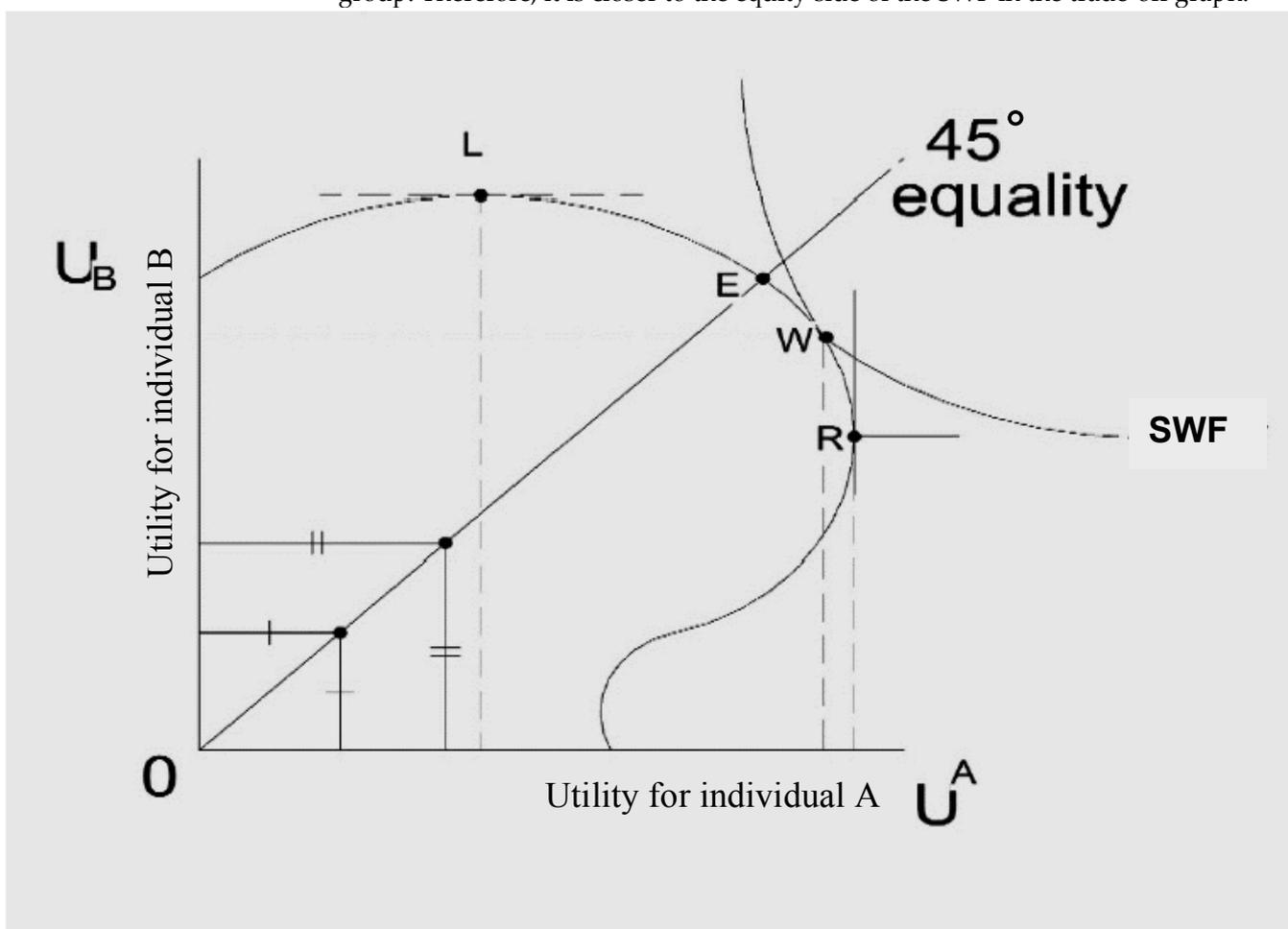

**Figure 1.** Position of the ethical theories and the Social Welfare Function in a 2-individual utility plane. The letter L stands for libertarianism, E for egalitarianism, W for weighted utilitarianism and R for Rawlsianism. $U^A$ is the utility for individual A while $U_B$ is the utility for individual B. Source: compiled by authors.

In Figure 1, we sketch a curve that mimics a nonlinear utility relationship between two individuals and some corresponding "Social Welfare Function" (SWF). Its profile reminds of the "optimal production point" of private vs. public goods according to welfare economics, as discussed by Mitchell and Carson (1989) or Gradinaru (2013). $U^A$ is the utility for individual A (horizontal axis), while $U_B$ is the utility for individual B (vertical axis). The four ethical theories are positioned according to their efficiency and equity merits: L stands for libertarianism, E for egalitarianism, W for weighted utilitarianism and R for Rawlsianism. The graph illustrates the trade-off between efficiency and equity in any society: libertarianism (L) is close to the vertical axis, at the maximum of the "utility relationship" curve, thus reflecting "more efficiency" while Rawlsianism (R) is closer to the horizontal axis representing "more equity", within some unavoidable inequality. An "optimal result" occurs at the tangency point between both curves, giving in this example more weight to A than B in the weighted utilitarianism approach as if A was among the worse-off people and B in the (more efficient) better-off people.

*7.2. UK Application*

As a convincing illustration, through the SWF graph, this analysis can be applied to UK government policy. The policies are simulated between the different ethical theories as explained sequentially below. From 1967 to 2003, the rate was changed five times; from 2003 to 2019, the rate remained constant. In 1989 and before, the UK used to apply the SOC method. From 1990 until the date of this paper, the STPR method has been applied in the UK.

Public policy in 1967 and 1969

The social discount rate was 8% in 1967 and 10% in 1969. These rates were set by the central government in real terms using the SOC approach. Although Spackman (1991) claims that before 1978 there was no clear procedure for setting the SDR, Pearce and Ulph (1995) have identified a rationale for setting these rates. According to them, the rationale for setting high SDR rates was the opportunity cost in terms of the marginal private sector investment forgone (Pearce and Ulph 1995). By that time, the required rate of return for public sector commercial companies was 8%.

In the trade-off curve, the rate of 8% represents high efficiency and minimum equity. High efficiency and low equity are features of a libertarian society. In such a society, a high-interest rate of 8% privileges the current generation. It encourages the current generation to take and consume as many resources as they please with no regard to future generations. In libertarianism theory, all generations are treated the same. Libertarianism does not differentiate between better-off and worse-off groups; all are free to reap the benefits of the open market. As a result, the people who already have power will be richer, and the lower classes are left to face their destiny empty-handed. However, from the point of view of other ethical theories, such as Rawlsian and weighted utilitarianism, it is important to distinguish between better off and the worse-off generations. Therefore, by analysing the 8% SDR from these two ethical theories perspectives, the following discussion points can be drawn.

When considering the Rawlsian and weighted utilitarian theories, two scenarios can be put in place. The first is the assumption that future generations are worse off than current generations. In this case, the public policy of applying a high SDR of 8% in 1967 cannot be a Rawlsian society because Rawlsian society assigns high weights to the worse-off people. In 1969, the SDR rate was increased again to 10%. This puts future generations in a worse position than they had been under the public policy of 1967. These high rates contradict even with the veil of ignorance and reciprocity principles in the Rawlsian school of thought. Likewise, the weighted utilitarian model seeks equality between all people but gives more weight to worse-off people. Thus, in the assumed scenario, the weighted utilitarian model also can be excluded, as it cannot be used as a justification for the high rates. Taking on many projects at high rates of 8 or 10% might have several consequences. High rates mean many resources are consumed today rather than tomorrow. It also means costs are transferred to future generations. The public policy of high discount rates is criticised by Broome (2008), who points out that current generations will not be affected by the consequences of their decisions, but future generations will pay and will be affected by the consequences of decisions taken today.

In the second scenario, the high rates can also be justified by Rawlsian theory if it is assumed that the current generation is worse off, even though it is hard to ascertain which generation is worse off. According to most traditional economic scenarios, there will be more prosperity and economic growth in future, and this will put future generations in a better position. Assuming that public policy was set on an ethical basis, according to the Rawlsian school of thought, the high rates of 8% and 10% are fair, bearing in mind that current generations are worse off. The impact of the weighted utilitarian theory in the second scenario, however, is indirect. Even if the present generation is held to be worse off, high rates of 8% and 10% cannot be justified by utilitarian theory because this theory seeks equality, with slightly higher weight given to the worse-off generations. Hence, a weighted utilitarian society would apply moderate rates rather than high rates. High rates are extremely partial to one generation.

The high rates make libertarianism a more convenient way to justify the unfairness of wealth distribution among different generations. The arguments here are not relevant to egalitarianism, which treats all groups exactly the same and where the trade-off between equity and efficiency will be zero. Weighted utilitarianism also attempts to equalise the two sides of the curve, but it differs from egalitarianism by assigning more weight to worse-off people. Thus, the public policies applied in 1967 and 1969 cannot be justified by this theory, taking into account that there will be continuous economic growth in the future scenario. On the other hand, if future generations will be worse off, as assumed above, then libertarianism is the perfectly suitable theory.

The high SDRs applied between 1967 to 1969 can be vindicated by the high economic growth the UK witnessed during the 1960s.

1978 and 1989 policies

The SDR was reduced in real terms to 5% in 1978 and 6% in 1989. The rate in 1978 was not imposed by the government; rather, it was left to market forces but with some agreement of the departments concerned (Spackman 1991, p. 2). It was a formal step to liberalise the market. Yet, the discount rate used in practice was 5% or slightly higher. By 1989 the increase in the discount rate was due to the rise in private investments in the UK during the 1970s. The discount rate of 6% was meant to reflect the cost of capital.

The reduction in the discount rate in 1978 and 1989 (compared to 8% SDR in 1967) was intended to shift the efficiency–equity trade-off curve toward the equity side. However, there was no evidence that it was designed purposely to meet this target. The SDR might have been higher in both years if the private sector rate of return had increased to more than the average of 5%.

Among all the public policies of setting SDR in the UK reviewed in this section, 5–6% could be considered as moderate rates. In the SWF graph, Figure 1, the 5 and 6% rates are located between the libertarian and egalitarian states. Theoretically, the 5% might reflect a more egalitarian society, but it did not really reflect the de facto liberal democracy economic system in the UK. The SDR was not reconciled with the whole economic system; rather, it was set as already mentioned solely in order to compensate for the cost of capital and the required rate of return (Spackman 1991). Whence, this proves that the UK public policies were isolated from any ethical considerations.

2003 SDR policy

In his guidance to determine the most appropriate SDR, Spackman attempted to address three points: "time preference rate, costing requires a cost of capital, and financial target setting for the enterprises requires an average return on total assets" (Spackman 1991, p. 3). Yet, Spackman argues that these three objectives cannot be covered at a single rate. Thus, the SDR in the UK most probably reflects only the time preference criteria and the cost of capital. Pearce and Ulph (1995) argue that there is no single appropriate SDR, but SDRs equal to 5% and 6% are too high. In 2003, the UK government reduced the SDR to 3.5%. It also changed the SDR approach from the SOC to the STPR. As discussed earlier, the STPR parameters were not based on ethical grounds but on economic factors. However, if the 3.5% rate is positioned in the trade-off curve, then 3.5% is found between the utilitarian and Rawlsian approaches. At least, this low 3.5% rate means that fewer resources are consumed by the current generations.

## 8. Conclusions

In the preceding sections, we have discussed the intergenerational equity issues caused by Public–Private Partnerships (PPPs). All the issues so raised are relevant to the SDR used in these Public–Private Partnerships projects. Therefore, the public policies of setting an appropriate SDR in PPPs do much matter and have been discussed. Several reasons are recalled and outlined.

First, public policymakers can manipulate the SDR used in Public Sector Comparator to make PPPs a better option in comparison to the traditional form of financing. This contradicts the VfM principle, which is the crux of PPPs. The manipulation of the SDR in the selection of financing method has two outcomes: (i) it proves that PPPs are the only option for the government, as already found by Grout (1997); (ii) it encourages policymakers to finance more public projects through PPPs. The second outcome has more impact on future generations. It allows the government to accumulate more debt from the private sector to deliver public projects. The increased level of debt will expose future generations to a fiscal crisis. These debts will be repaid in future through taxes or government spending cuts. To avoid this issue, the government is supposed to track the mounting debts from PPPs and measure the impact of these debts on future generations, yet there is no evidence that government uses sophisticated systems to track and measure the impact of these accumulated debts on its fiscal policy. PPPs debts are off-balance sheet items and, nowadays, are constraints on development's necessities. Thus, the policymakers should look at least at the impact of those debts on the fiscal deficit and on their evolution. To achieve that, governments should implement sensitivity analyses and examine worst-case scenarios. There is also an urgent need for governments to install sophisticated but efficient systems to keep pace with those debts, ensure transparency and resilience.

Secondly, the 2003 decision to reduce the SDR in the UK was not taken for ethical reasons. It was reduced because the 6% rate used before 2003 in the UK was too high (Pearce and Ulph 1995). The method used to determine SDR before 2003 was SOC. The main focus of SOC was efficiency. Our study leads us to argue that if the PPP project's objectives are to offer social benefits to society, as in the case of PPPs, then STPR is its best option.

In addition, setting the appropriate SDR requires distinguishing between the worse-off generations and the better-off generations, yet there is no consensus among economists on which generations are better-off. Thus, the worse-off generation state cannot be wholly predicted because of the uncertainty factor. We argue (bis repetita placet) that policymakers are morally obliged to revisit ethical theories, namely libertarian, egalitarian, utilitarian and Rawlsian, as guidance for forming public policies with more considerations on the impact of these policies on both worse and better-off people. These theories oblige the policymakers to comply with their various merits, whence to be thinking on drafting public policies on an equitable basis. For example, we emphasise the veil of ignorance of the Rawsilian theory for a more equitable society. Hence, the policymakers are under pressure to position themselves within the worse-off social class when they issue new policies. The policymakers must choose to increase the overall society's wellbeings regardless of which social class these policies affect. In such cases, policymakers will draw on classic utilitarian theory merits. Public policies that are established on ethical merits would create alignment between a state eco-political system and moral considerations.

Third, we reveal that public policy is not harmonised with ethical theories. The high SDR before 2003 reflects the high growth in that period. However, it also proves that the reduction in SDR is not based on the equitable distribution of resources between generations. In fact, we notice that various other considerations might be taken into account in setting the SDR, such as climate change. For example, recall that the UK SDR was so reduced to 3.5% in 2003. In 2003, the UK also changed the approach from SOC to STPR to take social considerations into account.

Fourth, more specifically, UK public policy was influenced by several ethical theories, such as libertarianism, egalitarian, weighted utilitarianism and Rawlsianism. The SDR used in the UK from 1967 to 2003 positions UK society in the trade-off curve. Throughout the analysis, it is assumed that each public policy positions a society within one of the ethical theories. For instance, the SDR was 8% in 1967 and 10% in 1969. These high rates fit a liberal society where the distribution of resources is allocated through market power. The high rates allocate more resources to the present generation. This feature is a feature of liberal society, where capitalists reap as much as they desire from the resources available with scant consideration for the worse-off. The social discount rate was reduced from 8% and 10% in 1967 and 1969 to 5% and 6% in 1978 and 1989, respectively. That locates UK society somewhere between libertarian and egalitarian. An egalitarian society treats all groups exactly the same, regardless of which group is worse off. By 2003, the SDR in the UK had been reduced to 3.5%. This SDR is somewhere between weighted utilitarian and Rawlsian societies in the trade-off curve. The weighted utilitarian model treats all generations the same. It is concerned with the aggregate wellbeing of the society, but more weight is allocated to the worse-off generation, whereas the Rawlsian model assumes that the total wellbeing of any society can be increased by increasing the wellbeing of the worst-off.

We conclude that public policies in the context of this paper could be strong monitoring of the PPPs debt levels and setting the appropriate SDR.

The findings in this report are subject to at least three limitations. First, due to practical constraints, this paper cannot provide a comprehensive review of the actual process of selecting PPPs projects. Hence, one cannot be definite about the decision-making process during the selection stage of PPPs. Investigating this stage is essential to know the quantitative measures taken into account in selecting projects other than the PSC; further research is needed in this area. Second, this study did not apply the suggested analytical framework of forming public policies to a genuine life case, but that seems much outside the present paper size (and aims). Third, this study did not measure the impact of marginal increase/decrease of the SDR on the intergenerational equity issues. Measuring this impact is an incentive for the worldwide government to decide on the appropriate value of SDR. Quantitative research could be interestingly conducted to address this issue.

**Author Contributions:** Conceptualization, AAY and MA; methodology, AAY; writing—original draft preparation, AAY; writing—review and editing, MA; visualization, AAY; supervision, MA. All authors have read and agreed to the published version of the manuscript.

**Funding:** MA has been partially supported by a grant of the Romanian National Authority for Scientific Research and Innovation, CNDS-UEFISCDI, project number PN-III-P4-IDPCCF-2016-0084.

**Institutional Review Board Statement:** not applicable

**Informed Consent Statement:** not applicable

**Data Availability Statement:** not applicable

**Conflicts of Interest:** The authors declare no conflict of interest.

**Glossary**

List of acronyms in alphabetical order and position of the first mention in the text (not necessarily the definition).

| | | |
|---|---|---|
| DMU | Diminishing Marginal Utility | 5.3 |
| GDP | Gross Domestic Product | 3.3.1 |
| GFS | Government Finance Statistics | 3.3.1 |
| NPV | Net Present Value | 3.2 |
| MCF | Marginal Cost of Funds Criterion | 2.1.1 |
| PIF | Private Finance Initiative | 1 |
| PPP | Public–Private Partnership | 1 |
| PSC | Public Sector Comparator | 3.2 |
| SDR | Social Discount Rate | 1 |
| SNA | System of National Accounts | 3.3.1 |
| SOC | Social Opportunity Cost | 1 |
| SPC | Shadow Price of Capital | 2.1.1 |
| STPR | Social Time Preference Rate | 1 |
| SWF | Social Welfare Function | 3.1 |
| UK | United Kingdom | 1 |
| VfM | Value for Money | 1 |